# Thermo-optic refraction in MoS₂ medium for "Normally On" all Optical switch

Pritam P Shetty[1], Mahalingam Babu[1], Dmitrii N Maksimov[2,3], Jayachandra Bingi[1, *]

[1] Bio-inspired Research and Development (BiRD) Laboratory, Photonic Devices and Sensors (PDS) Laboratory, Indian Institute of Information Technology Design and Manufacturing (IIITDM), Kancheepuram, Chennai 600127, India

[2] Kirensky Institute of Physics, Federal Research Center KSC SB RAS, 660036 Krasnoyarsk, Russia

[3] Siberian Federal University, Krasnoyarsk 660041, Russia

* Corresponding author, email-id: bingi@iiitdm.ac.in

## Abstract

Two dimensional (2D) nanomaterials like Molybdenum disulfide (MoS₂) have been drawing a lot of interest due to their excellent nonlinear optical response. In this research we study thermal lens formation in MoS₂ nanoflakes dispersion using mode mismatched pump probe configuration. Observation of the pump and probe beam intensity patterns gave visual insights on time evolution of photothermal lens formation. Effect of MoS₂ nanoflakes concentration on thermo-optic properties of dispersions were studied using thermal lens spectroscopy technique. Further, a thermo-optic refraction based technique to measure thermal lens size is proposed. Thermal lens region size increased with increase in pump power. The observed thermal lens modulation is applied to demonstrate 'normally on' all optical switch which showed excellent modulation of output beam signal by pump beam.

## Introduction

Molybdenum disulfide (MoS₂) is a 2D nanomaterial and lately has been under lot of attention by research community. A monolayer MoS₂ has two planes of hexagonally arranged sulfur atoms and one plane of hexagonally arranged molybdenum atoms sandwiched between them[1]. MoS₂ is attractive due to its good thermodynamic[2], mechanical[3], optical properties[4][5] and semiconducting nature[6]. Owing to this, 2D MoS₂ finds its application in Biosensing[7], microelectronics[8], optoelectronics[9], catalysis[10], lubrication and non-linear optics[11][12]. Liquid phase exfoliation technique is widely used to produce high quality and quantity of MoS₂ nanoflakes solutions with each nanoflake comprising of one or few 2D layers of MoS₂[13].

The optical Kerr effect is a phenomenon observed in some materials when an intense beam of light passes through them. This causes a change in the refractive index of material proportional to the local intensity of light. Further, this effect is also responsible of nonlinear optical effects like self-focusing/defocusing and spatial self-phase modulation (SSPM). SSPM causes the emerging beam from that material to self-interfere and have a concentric ring like intensity pattern (diffraction rings). The optical Kerr effect is also observed in MoS₂ dispersions[14], [15]. Dynamics of SSPM has been studied theoretically by proposing different models or by observing changes in diffraction ring [15]–[17]. All optical switching using nanomaterial dispersions have been demonstrated, but output of these devices varies as a function of number of rings[18] or ring deformation[19]. This does not give a precise control over the output. We take the approach of thermo-optic refraction rather than diffraction for optical switch, this opens the possibility for scaling of the device to microlevel. Proposed device in micro-scale can have lower power consumption and higher response time. Srivastava et. al. have demonstrated $MoS_2$ based micro optical device for ultra-fast optical switching in different pump probe configuration[20]

In this research we probe the region of interaction of pump laser beam with $MoS_2$ dispersions using a second diverging beam (probe beam). This gives us a visual insight into the dynamics of photothermal lens formation within $MoS_2$ dispersion and its effect on temporal diffraction ring evolution of the pump beam. The effect of pump beam power on the thermal lens region size is studied. Thermal lens spectroscopy (TLS) technique is used to calculate thermo-optic properties of the dispersions with varying concentration of $MoS_2$ nanoflakes. Further, a 'normally on' all-optical switch is demonstrated.

## Results and Discussions

### Material preparation and characterization

$MoS_2$ powder (98% pure) is grinded by Mortar and Pestel for 5 hours using Methanol as solvent and dried for four days. 3%w/w Polyvinylpyrrolidone (PVP) polymer solution is prepared by mixing PVP with ethanol/acetonitrile blend as solvent. The solvent blend is prepared by mixing of ethanol and acetonitrile in ratio 1:1. $MoS_2$ dispersions were prepared by adding 2%w/w grinded $MoS_2$ to the PVP polymer solution. The PVP polymer acts as colloidal stabilizer. The dispersions are sonicated for three hours. After sonication, dispersions were centrifuged for two hours. Centrifugation allows bulk $MoS_2$ to settle at bottom and we can extract the colloidal solution with $MoS_2$ nanoflakes by pipetting. This freshly prepared $MoS_2$ dispersion is diluted by two-fold series dilution with addition of PVP solution thrice. These diluted samples are later used to study the changes in thermo-optic properties of dispersion as a function of the $MoS_2$ nanoflakes concentration.

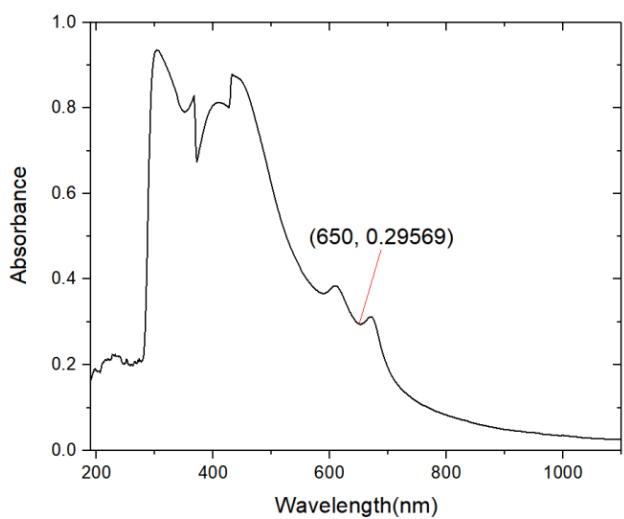

*Figure 1:Absorption spectrum of MoS₂ nanoflake dispersion.*

The UV-Vis spectrum as shown in figure 1 indicates that the dispersions have broadband absorption in visible region(400-700nm) and can be a prospective optical material in the visible region. In this research, light of 650nm wavelength is used with the absorption coefficient of 0.681cm$^{-1}$.

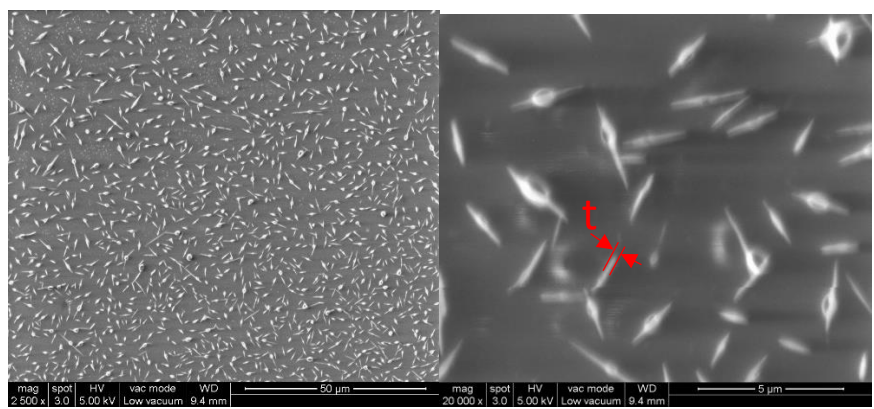

*Figure 2: SEM micrograph of thin films of MoS₂ nanoflakes dispersion on glass slide. (right)zoomed view.*

The dispersions were spin coated on glass slide for SEM analysis. As seen in figure 2 the nanoflakes had uniform size and average thickness of 153.5nm ± 23nm.

## Evolution of thermal lens region and Quantifying its size

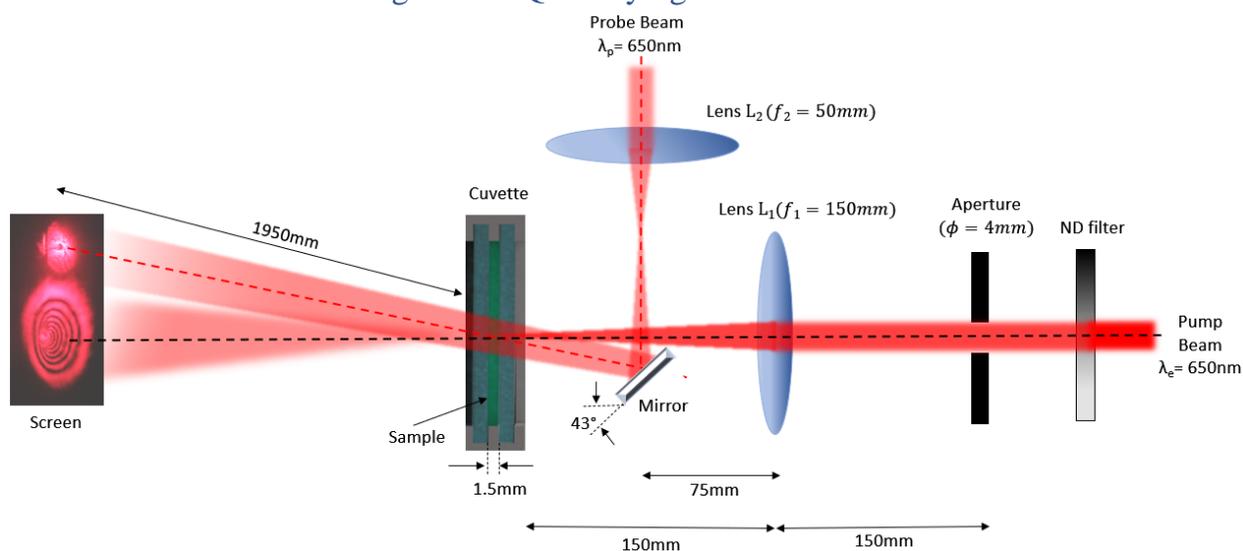

*Figure 3:Experimental setup to study time evolution and quantification of thermal lens region. (distance between L₂ to Beam splitter=18cm)*

The MoS₂ dispersion samples were poured into a glass cuvette of path length 1.5mm. A collimated pump beam generated by a laser diode ($\lambda e = 650nm$) is focused on the sample using lens $L_1$. The power of pump beam is controlled by a variable neutral density (ND) filter. A collimated beam from diode laser ($\lambda p = 650nm$) is used for probe beam. Using lens $L_2$, probe beam of divergence 67mrad is passed through focal point of the pump beam on the sample to study thermal lensing effect. The probe beam is passed through the sample at a small angle of ~2° with respect to the axis of the pump beam to separate the pump and probe intensity pattern on the screen.

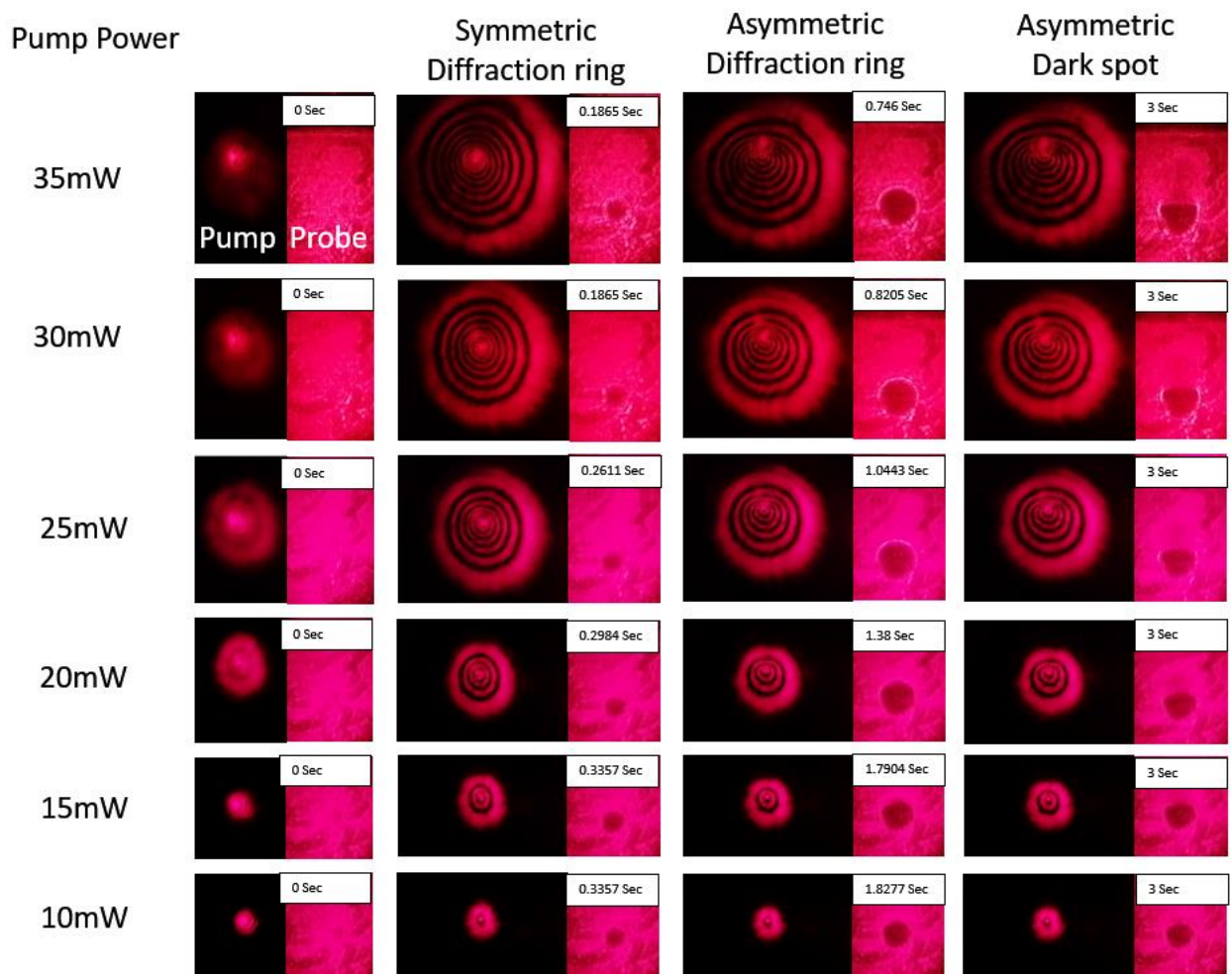

*Figure 4: Time evolution of diffraction rings and thermal lens dark spot for different pump power(inset: Elapsed time). See supplementary material for video.*

When the pump beam is focused on $MoS_2$ dispersion, there is intense localized heat generation due to absorption by suspended nanoflakes. Three-dimensional thermal distribution at the focal point inside the sample can be approximated to have spherical gradient (assuming there is no convective heat flow in dispersion). This causes a corresponding spherical gradient refractive index change in the $MoS_2$ dispersion with a very low refractive index at the center. Due to intense radial refractive index change encountered by pump beam, it undergoes self-interference along the direction of propagation causing concentric diffraction rings on the screen as shown in figure 4. There are three stages found in time evolution of the thermal lens region when the pump beam and probe beam output from the sample is observed-

*Symmetric diffraction rings*: The intensity distribution of probe beam forms a symmetric diffraction beam. This indicates there is no convective heat flow and a perfect spherical gradient refractive index distribution is formed at the pump beam waist. A tiny circular dark spot can be seen in the probe beam (figure 4). Strong localized heating by the pump beam causes a spherical gradient refractive index lens with low refractive index at the center and increasing along the radial distance. This thermal lens strongly diverges the part of probe beam encountered by it, causing a dark spot.

*Asymmetric diffraction rings*: As the pump beam waist radius(18.9µm) is far smaller than the thermal lens region(~1mm), the intensity distribution of the pump beam (diffraction rings) is very sensitive to any convective heat flow in dispersions. Hence when convective heat flow emerges, it is first indicated by collapsing of the diffraction ring in the pump beam but the dark spot in probe beam still appears circular.

*Asymmetric dark spot*: When equilibrium is reached between heat carried by convection and heat generation by the pump beam, a stable asymmetric dark spot can be seen in the probe beam. This dark spot in the probe beam along with the thermal plume due to convection can be clearly seen in figure 5. In figure 4 the intensity patterns of the asymmetric dark spot for all pump powers is taken at elapsed time of 3 seconds and does not represent the time of equilibrium. However, it is observed that the asymmetric dark spot formation is faster at higher pump powers.

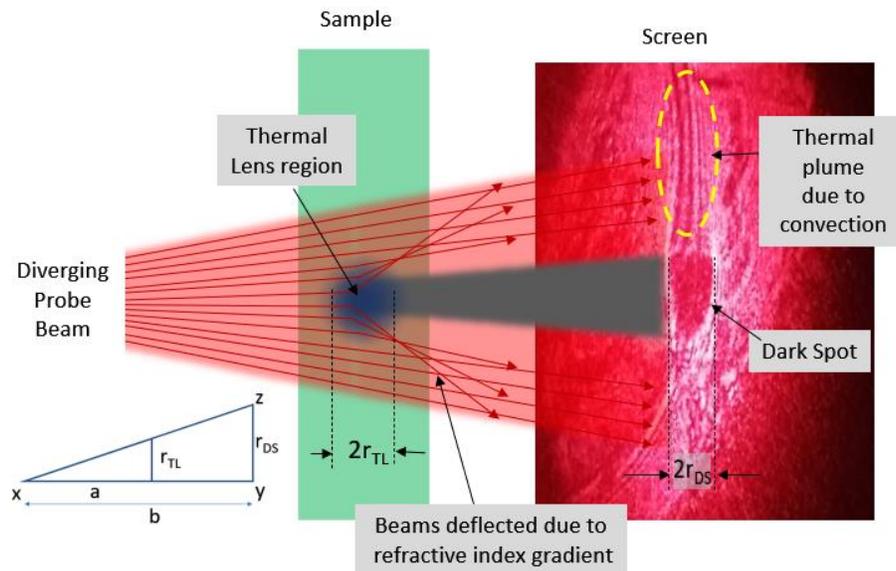

*Figure 5: Light rays from the probe beam which encounter thermal lens region get deflected due to refractive index gradient*

By knowing the size of the dark spot, we can approximately calculate the diameter of the thermal lens region. It is assumed that there is no deflection of rays at the edge of the thermal lens region. Hence the probe rays emerging from the focal point of lens L$_2$ subtends an angle $\angle zxy$ with distances r$_{TL}$ and r$_{DS}$ respectively (figure 5 inset). Here, r$_{TL}$ and r$_{DS}$ are the radius of the thermal lens region within the sample and the radius of the dark spot on the screen, respectively. Here r$_{DS}$ is measured horizontally on the screen as dimensions in this direction is not affected by the convective flow in the dispersion. By using a simple relation, we can find the radius of the thermal lens region:

$$r_{TL} = \frac{a}{b} * r_{DS}$$

Time evolution of the thermal lens region and thermal lens size is recorded by varying the pump power from 5 to 35mW as shown in figure 4 and table 1.

*Table 1: Variation of thermal lens size with pump power.*

| Pump power(mW) | $r_{ds}$ (mm) | $r_{TL}$ (mm) |
|---|---|---|
| 35 | 10.145 | 0.96507 |
| 30 | 9.5 | 0.903712 |
| 25 | 9.3 | 0.884687 |
| 20 | 8.8 | 0.837123 |
| 15 | 8.3 | 0.789559 |
| 10 | 7.5 | 0.713457 |
| 5 | 6.25 | 0.594548 |

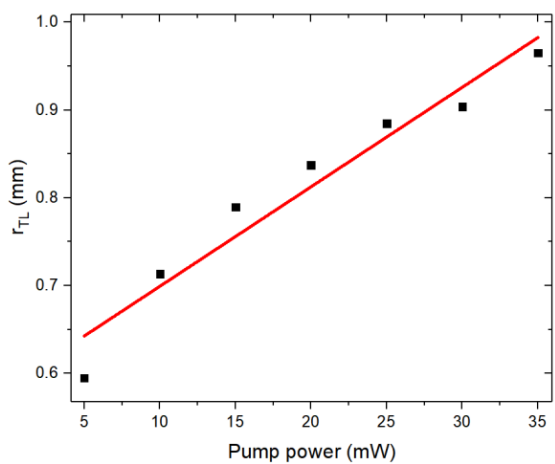

*Figure 6: Dependence of the thermal lens size on the pump power.*

It can be observed from figure 4 that as the pump power is increased the time taken to form the asymmetric diffraction ring reduces i.e. the convection flow is stronger. At pump power less than 10mW, the diffraction rings as well as the dark spot remain symmetric signifying there is very low thermal convection. The number of diffraction rings in the pump beam reduces with reduction in the pump power showing that the radial thermal gradient is also reduced. This is confirmed in table 1 where the thermal lens size is reduced by 38.4% as the pump power is reduced from 35mW to 5mW. Figure 6 shows a linear dependence of the thermal lens size on the pump power. Hence, for micro-optical devices with few layers of $MoS_2$, the pump power requirement will be low. This will be ideal for low power device application.

## Thermal lens spectroscopic (TLS) analysis of MoS$_2$ dispersion

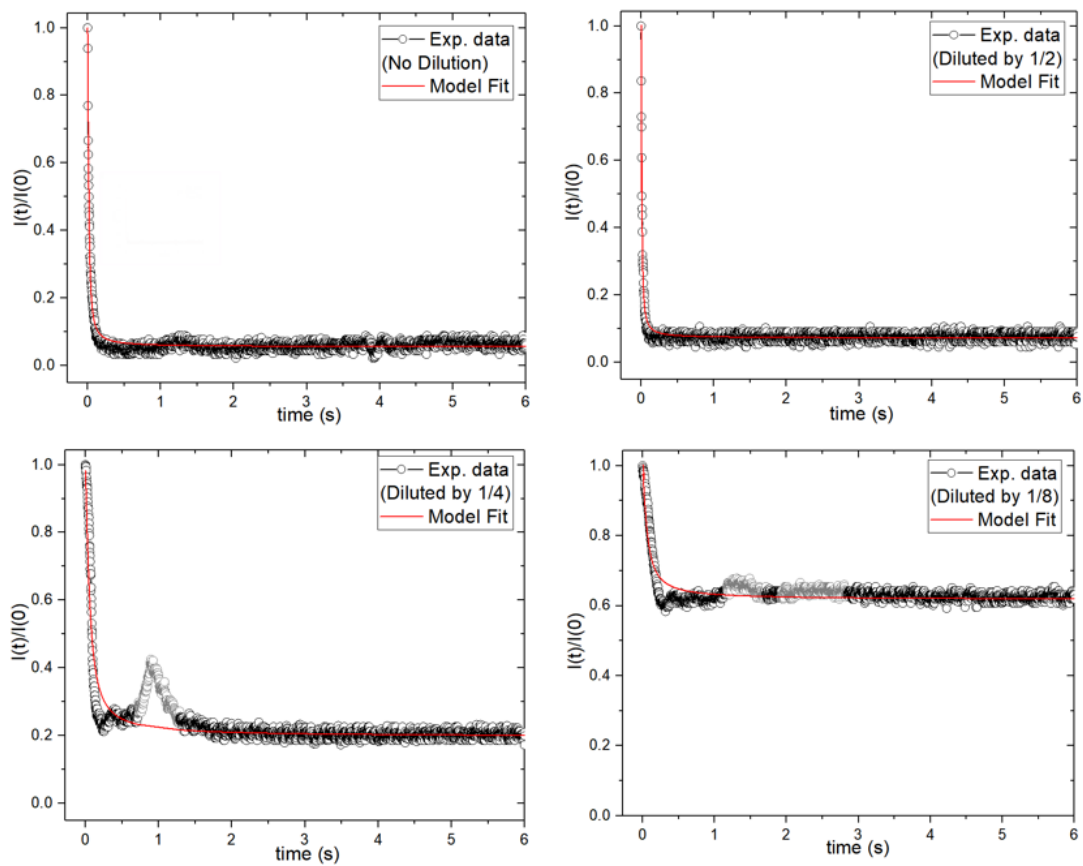

*Figure 7: Thermal lens signal variation with time for different dilution of the sample (greyed out sample points are not considered for model fit)*

The MoS$_2$ dispersion were analyzed by TLS method in dual beam mode mismatched configuration as shown in the experimental setup in figure 3. To record the thermal lens (TL) signal I(t), a photodiode connected to an oscilloscope is placed at the center of the dark spot in the probe beam transmitted through sample. By fitting an analytical expression based on the theoretical model by J Shen et al [21] to experimental TL signal data, various thermal and optical parameters of the MoS$_2$ dispersion can be calculated. The model is developed using Fresnel diffraction theory. Here we calculate the temperature co-efficient of refractive index (dn/dT) for the MoS$_2$ dispersion samples which are diluted by series two-fold dilution three times. According to the model, the temporal change in the TL signal is given by

$$I(t) = I(0) \left\{ 1 - \frac{\theta}{2} \arctan \left[ \frac{2mV}{[(1+2m)^2+V^2]\left(\frac{t_c}{2t}\right)+1+2m+V^2} \right] \right\}^2 \quad \ldots\ldots\ldots (1)$$

Here,

$I(0)$- intensity of the TL signal when either t or θ is zero

$\omega_{0p}$ – probe beam waist radius=6.31μm

$\omega_{1p}$ – probe beam radius at the sample=5.75mm

$\omega_e$ – pump beam waist radius=18.93μm

$$V = \frac{Z_1}{Z_c} = \frac{\textit{Probe beam waist to sample distance}}{\frac{\pi \omega_{0p}^2}{\lambda_p}} = \frac{21.5 * 10^{-2}}{1.9244 * 10^{-4}} = 1117.23$$

$$m = \left(\frac{\omega_{1p}}{\omega_e}\right)^2 = 92264.46$$

After fitting equation (1) to experimental TL signal data, we can get values of constants $t_c$ and $\theta$. Parameter $t_c$ is a characteristic time constant which depends on properties of dispersion like specific heat and density. Parameter $\theta$ is a dimensionless variable which depends on strength of thermal lens. It is approximately equal to thermally induced phase shift of probe beam between center and edge of dark spot at steady state of thermal lens. Further, we can calculate the temperature coefficient of the refractive index, $\frac{dn}{dT} = \frac{-\theta K \lambda_p}{PAL\varphi}$.

Here,

K - thermal conductivity of the dispersion = ~0.1769W/mK[22]

P - pump beam (excitation) power = 30mW

A - absorption co-efficient of the dispersions= 68.097m$^{-1}$

L- sample thickness= 1.5mm

φ- fraction of excitation energy converted into heat=1

Negative sign of dn/dT is due to negative photothermal lens generated by the pump beam within the sample. Assuming that the thermal conductivity (K) and absorption co-efficient (A) are not changing significantly with concentration of MoS$_2$ nanoflakes, dn/dT is calculated. The trend of dn/dT is shown in figure 8.

*Table 2: Constants $t_c$ and $\theta$ extracted by fitting the model to thermal lens experimental data as shown in figure 7 for different dilution of MoS$_2$ dispersions.*

| Dilution | $t_c$ (msec) | $\theta$ |
|---|---|---|
| 0 | 0.29 | 0.9774 |
| 1/2 | 0.1767 | 0.9348 |
| 1/4 | 0.9494 | 0.7109 |
| 1/8 | 0.8568 | 0.27393 |

With dilution higher than 1/2, small peaks are observed in the TL signal just after formation of the thermal lens. These peaks are due to perturbation in dispersions. A possible explanation for this is that with higher dilution, the concentration of the MoS$_2$ nanoflakes is lower and there is a weak thermal lens formation. Hence the photothermal lens formed gets easily distorted by weak convective flows. Since the TL signal has already reached lowest point, datapoints in these peaks (greyed out data points in figure 7) are masked and not considered for the model fit. Table 2 shows the values of constants $t_c$ and $\theta$ extracted from the experimental data by fitting to the model as shown in figure 7.

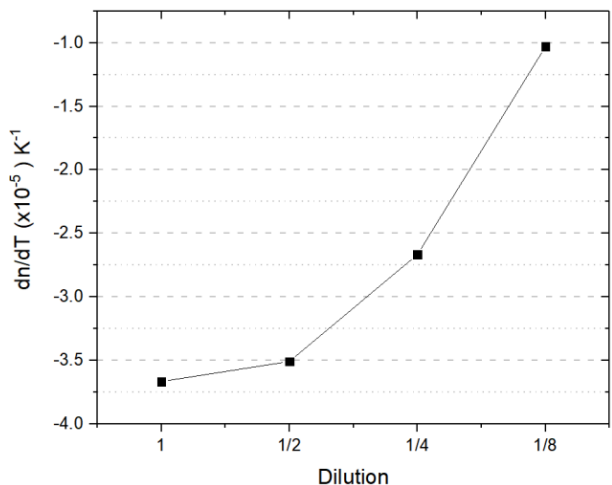

*Figure 8: Change in thermal co-efficient of refractive index(dn/dT) for two-fold series dilution of MoS₂ dispersions*

Temperature dependent change in the refractive index (dn/dT) is found to decrease with increase in dilution of the MoS₂ dispersion (Figure 8). This dependency can be explained as follows: large surface area of MoS₂ nanoflakes allows higher absorption of light as well as efficient transfer of heat to other molecules. At higher concentration of MoS₂ nanoflakes (i.e. at lower dilution of dispersion), there is greater absorption of the pump beam and more nanoflakes act as a heat source. Further, this heat gets absorbed by the neighboring solvent molecules causing large refractive index gradient. Thermo-optic coefficient(dn/dT) of nano layer MoS₂ is reported as $\sim 10^{-5} K^{-1}$ [23,24] . Whereas of common solvent(ethanol) is $\sim -10^{-4} K^{-1}$[25]. In this research thermo-optic coefficient of MoS₂ nanoflakes dispersion is measured to be $\sim -3.66 \times 10^{-5} K^{-1}$ and decreases up to $\sim -1.03 \times 10^{-5} K^{-1}$ after dilution to 1/8[th] of initial concentration. This indicates the effective thermo-optic co-efficient of MoS₂ nanoflakes dispersion depends primarily on dn/dT of solvent and proportional to MoS₂ nanoflakes concentration. Hence thermo-optic properties of the MoS₂ dispersions can be optimized by changing the concentration of MoS₂ nanoflakes and the solvent.

## Application as "Normally ON" all optical switch

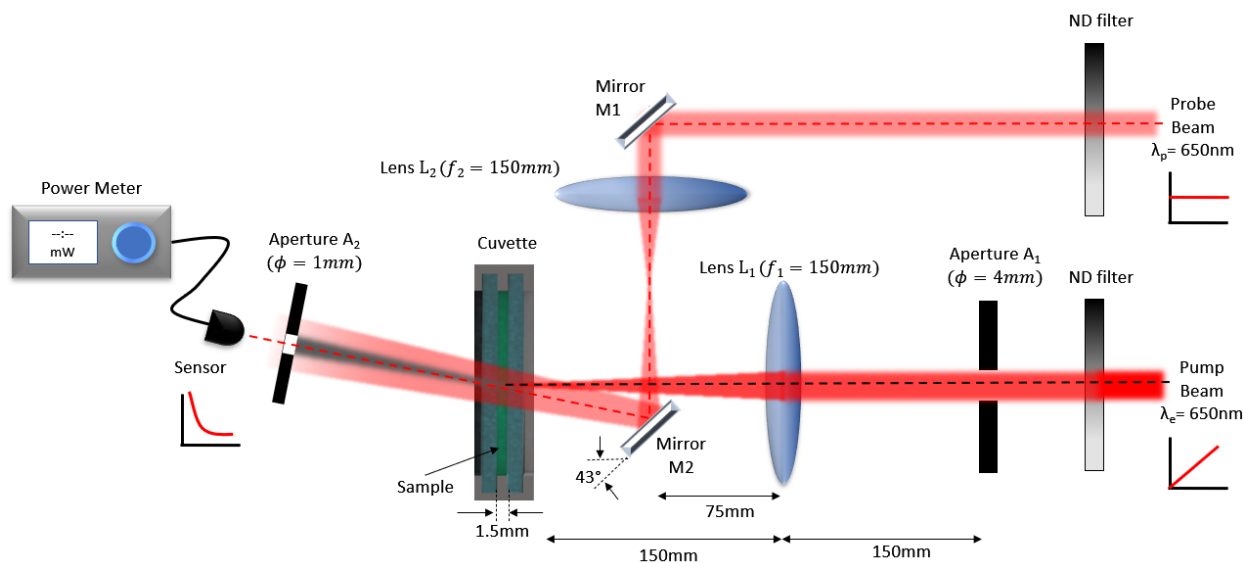

*Figure 9: Experimental setup for "normally on" all optical switch (dsitance between $L_2$ and $M_2$=18cm, $A_2$ and sample=5cm). Pump beam emerging out of the sample inthe cuvette is not shown .*

Figure 9 shows the experimental setup to demonstrate $MoS_2$ dispersions as prospective material for the "normally on" all optical switch. The three terminals of the optical switch comprise of two input beams i.e. the pump and probe beams, and the part of the diverging probe beam transmitted through the thermal lens region in dispersion. The latter is considered as the output. To avoid heating of the proposed device, the wavelength of probe beam should be chosen to have least absorbance by the dispersion. The thermal lens is created by focusing the pump beam using lens $L_1$. Beam powers are controlled by using a variable Neutral Density (ND) filter and measured using PM320E Thorlabs optical power meter with a Si photodiode as a sensor. By keeping the probe power constant, the output power is measured by the varying pump power from 0 to 70mW in step of 5mW increment. Measurements are repeated for four different probe powers from 5 to 25mw.

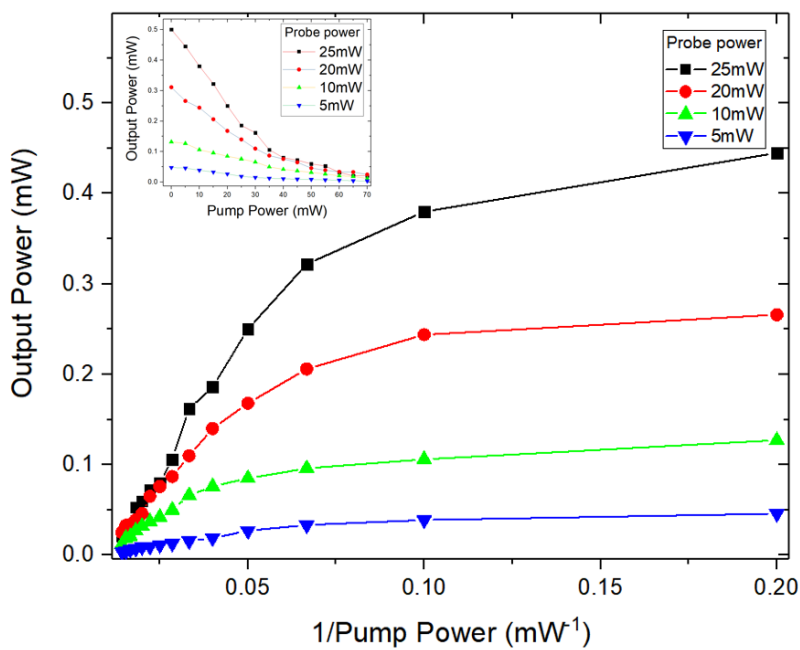

*Figure 10:Output characteristics of the "normally on" all optical switch shows output power variation vs 1/(input pump power), inset- output power vs input pump power*

Figure 10 shows output characteristics of the thermo-optic refraction based all optical switch. This graph looks very similar to the output characteristics of the conventional electronic transistor. The output switch-off time for the probe power of 25mW and pump power of 30mW is found to be ~0.16 sec. We deduce following conclusions from figure 10(inset): the amplitude of the output power when the pump power is zero can be controlled by the probe beam power. The slope of the linear response in the output power below 15mw of the pump power can be increased by increasing the probe power. Therefore, there is good control over modulation of output signal in thermo-optic refraction based approach for optical switching.

The present configuration contains $MoS_2$ in colloidal form where there is a limitation in response time due to solvent. In real life application the demonstrated concept (thermo-optic refraction in $MoS_2$) in this research should be applied as a single/few layer $MoS_2$ (solid state) sandwiched between micro laser(pump) and substrate. A second micro laser orthogonal to this arrangement can act as probe. In this micro device one can expect better parameters liker higher switching speed and low power.

## Conclusion

In this work we visually studied the time evolution of the thermal lens (TL) region in $MoS_2$ dispersions by pump-probe configuration. It is visually confirmed that there is a collapse of diffraction rings of the transmitted pump beam due to a vertical convective heat flow. Size of the TL region is reduced by 38.4% when the pump power is varied from 35 to 5 mW. The thermo-optic properties of $MoS_2$ dispersions were studied using TLS technique. It is found that the temperature coefficient of the refractive index increases with increase in concentration of $MoS_2$ nanoflakes in the dispersion. By considering thermo-optic refraction as a tool for optical modulation, a 'normally on' all optical switch is demonstrated using $MoS_2$ dispersions.

There is an excellent modulation of output beam using the pump beam. Low power micro-optical devices based on single monolayer $MoS_2$ can be explored in future work.

## Author contributions

**Pritam P Shetty**: Conceptualization, Methodology, Validation, Investigation, Visualization, Writing - Original Draft **Mahalingam Babu**: Investigation **Dmitrii N Maksimov**: Writing - Review & Editing **Jayachandra Bingi**: Conceptualization, Methodology, Writing - Review & Editing, Supervision.

## Declaration of Competing Interest

The authors declare that they have no known competing financial interests or personal relationships that could have appeared to influence the work reported in this paper.

## Acknowledgements

Authors acknowledge the funding support from DST India under INT/RUS/RFBR/P-262.